\documentclass[12pt]{article}

\usepackage{amsmath}
\usepackage{amssymb}

\usepackage{cite}

\begin{document}

\begin{center}
{\bf Conservation Laws of One-Dimensional Equations 
\\
of Relativistic Gas Dynamics in Lagrangian Coordinates}

\vspace*{3mm}

{Warisa Nakpim and Sergey V. Meleshko}%

\vspace*{3mm}

{School of Mathematics, Faculty of Science, 
\\
Khon Kaen University, Khon Kaen, 40002, Thailand}

{School of Mathematics, Institute of Science, Suranaree University of
Technology, 
\\
Nakhon Ratchasima, 30000, Thailand}

\end{center}

\begin{abstract}
The present paper is focused on the analysis of the one-dimensional
relativistic gas dynamics equations. The studied equations are considered
in Lagrangian description, making it possible to find a Lagrangian
such that the relativistic gas dynamics equations can be rewritten
in a variational form. Complete group analysis of the Euler-Lagrange
equation is performed. The symmetries found are used to derive conservation
laws in Lagrangian variables by means of Noether's theorem. The analogs
of the newly found conservation laws in Eulerian coordinates are presented
as well.
\end{abstract}

\section{Introduction}

Conservation laws play a significant role in physics: they describe
essential properties of the processes modeled by a given system of
partial differential equations. Conservation laws given by densities
of local form determine physically important conserved quantities
such as energy, momentum, angular momentum, mass, charge, etc. which
are constants of motion central to an analysis of the time evolution
of the fields. 
A conservation law is a divergence expression that vanishes on all
solutions of the system. The most widely-known tool for finding conservation
laws of differential equations is Noether's theorem \cite{bk:Noether[1918]}.
It can be applied, in principle, to any system of differential equations,
that admits a variational formulation in terms of a Lagrangian. Hence,
the problem of finding a variational principle for a given system
of equations arises naturally. A variational principle constitutes
an alternative method for determining the state or dynamics of a physical
system. Efforts in this area involve the construction of integral
functional encompassing all the equations of a specific problem.

Among the studies devoted to the variational principles of non-re\-la\-tivistic
gas dynamics, one can distinguish the following approaches \cite{bk:Shmyglevski}.
In the approach \cite{bk:Manwell} the stationary stream functions
of plane flows were used. In \cite{bk:Kraiko1981} it was noted that
the functional used in \cite{bk:Manwell} had already been introduced
in \cite{bk:LinRubinov}, and using two stream functions, the author
of \cite{bk:Kraiko1981} proposed a similar variational principle
for stationary three-dimensional gas flows. For potential flow there
is also a Lagrangian such that the gas dynamics equations can be obtained
from the variational principle. Another approach was started by the
author of \cite{bk:Bateman}, where a Lagrangian was constructed
by using Lagrangian multipliers. Later this approach was also applied
to magneto-hydrodynamics \cite{bk:Shmyglevski}. As noted in \cite{bk:SeligerWhitham[1968]},
the Eulerian description raises mathematical problems for finding
a variational formulation of the studied problem. On the other hand,
the study of the non-relativistic gas dynamics equations in Lagrangian
coordinates allows one to use the natural Lagrangian consisting of
the potential and kinetic energies \cite{bk:SeligerWhitham[1968],bk:Webb2018,bk:SiriwatKaewmaneeMeleshko2016,bk:DorodnitsynKozlovMeleshko2019}\footnote{See also references therein.}

Various works have been proposed \cite{Taub_2011,bk:Schutz[1970],Elze_1999,bk:MonaghanPrice_2001,Ootsuka_2016}
to construct Lagrangian formulation of a relativistic perfect fluid.
However, we have not yet uncounted a Lagrangian ${\cal {L}}$ in the
form of applied in the present paper. This Lagrangian allows us to
write the relativistic gas dynamics equations in Lagrangian description
$(\xi,t)$ as the Euler-Lagrange equation ${\displaystyle \frac{\delta{\cal {L}}}{\delta\varphi}=0}$,
where $(x,t)$ are Eulerian coordinates and $x=\varphi(\xi,t)$.

Noether's theorem relates symmetries and conservation laws, because
any variational or divergent symmetry is defined by an admitted generator.
Thus, for applying Noether's theorem one needs to know the admitted
Lie algebra. One of the tools for studying symmetries is the group
analysis method \cite{bk:Ovsiannikov1978,bk:Olver[1986]}, which
is a basic method for constructing exact solutions of partial differential
equations.

The present paper is focused on the one-dimensional relativistic gas
dynamics equations of a polytropic gas in Lagrangian description,
with the objective to analyze the variational second-order partial
differential equation to which the relativistic gas dynamics equations
are reduced in Lagrangian coordinates. Using Noether's theorem, conservation
laws in Lagrangian and Eulerian descriptions are found.

This paper is organized as follows. 
Section 2 introduces the relativistic gas dynamics equations and their
variational formulation. In Section 3 we consider the general case
and the corresponding conservation laws. Then, we present the cases
with additional symmetries and conservation laws in Section 4. Finally,
Section 5 gives concluding remarks.

\section{Relativistic gas dynamics equations}

In this section the relativistic one-dimensional equations of a perfect
fluid are considered in Eulerian and Lagrangian coordinates.

\subsection{The relativistic gas dynamics equations in Eulerian coordinates}

The motion of a relativistic perfect fluid is governed by the equations
expressing the conservation of the number density, and the conservation
of the energy and momentum \cite{bk:LandauLifshitz}, \cite{bk:Anile_1989}.
In the one-dimensional case these equations are
\begin{equation}
\begin{array}{l}
{\displaystyle \frac{\partial}{\partial t}(\Gamma^{-1}n)+{\displaystyle \frac{\partial}{\partial x}(\Gamma^{-1}nv)=0,}}\\[1.5ex]
{\displaystyle \frac{\partial}{\partial t}\left(T^{10}\right)+{\displaystyle \frac{\partial}{\partial x}\left(T^{11}\right)=0,}}\\[1.5ex]
{\displaystyle \frac{\partial}{\partial t}\left(T^{00}\right)+{\displaystyle \frac{\partial}{\partial x}\left(T^{01}\right)=0,}}
\end{array}\label{eq:Nov25.1}
\end{equation}
where
\begin{gather*}
{\displaystyle T^{00}=(e+p)\Gamma^{-2}-p,\quad T^{01}=T^{10}={\displaystyle \dfrac{v}{c}}(e+p)\Gamma^{-2},}\\
\quad T^{11}=\dfrac{v^{2}}{c^{2}}(e+p)\Gamma^{-2}p
\end{gather*}
are components of the energy-impulse tensor, $n$ is the proper number
density of particles, $e$ is the total energy density, $v$ is the
velocity, $c$ is the speed of light, ${\displaystyle \Gamma=\sqrt{1-v^{2}/c^{2}}}$,
and $p$ is the pressure. After scaling $t$, $x$ and $v$, one can
assume that $c=1$.

A perfect fluid is a two-parameter thermodynamical model, which means
that two of the thermodynamical variables $e$, $p$, $n$, $\eta$,
and $T$ define the remaining variables. Here $\theta$ is the absolute
temperature, and $\eta$ is the specific entropy. The thermodynamical
variables also satisfy the thermodynamical identity
\[
\theta d\eta=d\left(\frac{e}{n}\right)+pd\left(\frac{1}{n}\right).
\]

In this article, we study a relativistic polytropic gas. Such gas
is defined by the state equations
\[
\varepsilon=c_{v}\theta,\,\,\,\,p=Rn\theta,
\]
where $\varepsilon$ is the specific internal energy, $R$ is the
gas constant, $c_{v}$ is the specific heat capacity at constant volume.
For a polytropic gas one obtains that \cite{bk:Ovsiannikov[2003]}
\[
p=Sn^{\gamma},
\]
where ${\displaystyle \gamma=1+\frac{R}{c_{v}}}$, and $S=Re^{(\eta-\eta_{0})/c_{v}}$
with some constant $\eta_{0}$. As $R>0$ and $c_{v}>0$, then $\gamma>1$.
In this case
\[
e+p=n+{\displaystyle \frac{\gamma}{\gamma-1}Sn^{\gamma},}
\]
and equations (\ref{eq:Nov25.1}) become
\begin{equation}
\begin{array}{c}
m_{t}+(mv)_{x}=0,\\[1.5ex]
n(v_{t}+vv_{x})+\Gamma^{4}n^{\gamma-1}(\gamma Sn_{x}+nS_{x})\\
+{\displaystyle \frac{n^{\gamma}S}{\gamma-1}\left((1+v^{2}-\gamma v^{2})v_{t}+\gamma(2-\gamma)vv_{x}\right)=0,}\\[1.5ex]
S_{t}+vS_{x}=0,
\end{array}\label{eq:Eulerian}
\end{equation}
where $m=n\Gamma^{-1}$.

\subsection{Lagrangian coordinates}

Alternatively to the Eulerian description, there is the Lagrangian
description, where the particles are identified by the positions which
they occupy at some initial time. Typically, Lagrangian coordinates
are not applied in the description of fluid motion, as the Lagrangian
type of specification leads to a cumbersome analysis \cite{bk:Batchelor}.
However, in some special contexts the Lagrangian description is indeed
useful.

\subsubsection{Deriving the gas dynamics equations in Lagrangian coordinates}

The Eulerian $(x,t)$ and Lagrangian $(X,t)$ coordinates are connected
by the relation $x=\varphi(X,t)$, where the function $\varphi$ satisfies
the Cauchy problem
\[
\varphi_{t}(X,t)=v(\varphi(X,t),t),\ \ \varphi(X,0)=X.
\]
In the Lagrangian coordinates $(X,t)$ the first equation of (\ref{eq:Eulerian})
can be integrated
\[
m=\frac{m_{0}}{\varphi_{X}},
\]
where $m_{0}(X)$ is the arbitrary function of integration. Using
the change
\begin{equation}
\xi=\alpha(X),\label{eq:change}
\end{equation}
where $\alpha^{\ \prime}(X)=m_{0}(X)$, one obtains that
\[
\bar{m}(t,\xi)=\frac{1}{\bar{\varphi}_{\xi}(t,\xi)}.
\]
Here the functions $\bar{\varphi}(t,\xi)$ and $\varphi(t,X)$ are
related by the formula
\[
\bar{\varphi}(t,\alpha(X))=\varphi(t,X).
\]
Further the sign $\bar{}\ $ is omitted. In the Lagrangian coordinates
the change (\ref{eq:change}) defines an equivalence transformation:
this change simplifies the equations studied. The coordinates $(\xi,t)$
are called the mass Lagrangian coordinates \cite{bk:RozhdYanenko[1978]}.

In the Lagrangian coordinates $(\xi,t)$ the general solution of the
third equation of (\ref{eq:Eulerian}) is
\[
S=S_{0}(\xi),
\]
where the function $S_{0}(\xi)$ is the arbitrary function of integration.

From the relations
\begin{equation}
v(\varphi(\xi,t),t)=\varphi_{t}(\xi,t),\quad m(\varphi(\xi,t),t)={\displaystyle \frac{1}{\varphi_{\xi}(\xi,t)}},\label{eq:def_lagr_var}
\end{equation}
one finds that
\[
\begin{array}{c}
v_{t}=\varphi_{tt}-\varphi_{t}\varphi_{\xi}^{-1}\varphi_{t\xi},\,\,\,v_{x}=\varphi_{\xi}^{-1}\varphi_{t\xi},\,\,\,m_{x}=-\varphi_{\xi}^{3}\varphi_{\xi\xi},\\[1.5ex]
m_{t}=-\varphi_{\xi}^{-2}(\varphi_{\xi t}+\varphi_{t}\varphi_{\xi}^{-1}\varphi_{\xi\xi}).
\end{array}
\]
Substituting these derivatives into the second equation of (\ref{eq:Eulerian}),
it becomes
\begin{equation}
\begin{array}{c}
{\displaystyle \left(\frac{\gamma}{\gamma-1}\left(\Gamma^{2}(\gamma-1)-\gamma+2\right)S_{0}+\Gamma^{1-\gamma}\varphi_{\xi}^{\gamma-1}\right)}\varphi_{\xi}^{2}\varphi_{tt}\\[2ex]
-2\gamma\Gamma^{2}S_{0}\varphi_{t}\varphi_{\xi}\varphi_{t\xi}+\Gamma^{4}(\varphi_{\xi}S_{0}^{\prime}-\gamma S_{0}\varphi_{\xi\xi})=0.
\end{array}\label{eq:Nov25.10}
\end{equation}

\subsubsection{Variational formulation}

It is well-known that in the non-relativistic gas dynamics, there
exists a Lagrangian whose Euler-Lagrange equation is similar to the
equation corresponding the gas dynamics equations in Lagrangian coordinates
\cite{bk:Webb2018,bk:DorodnitsynKozlovMeleshko2019}, i.e., there
is a Lagrangian ${\cal L}$ such that this equation is equivalent
to the equation ${\displaystyle \frac{\delta{\cal L}}{\delta\varphi}=0}$.
This makes it possible to assume that equation (\ref{eq:Nov25.10})
also has this very property. Here ${\displaystyle \frac{\delta}{\delta\varphi}}$
is the variational derivative.

Recall that the Euler-Lagrange equation is computed as
\begin{equation}
\frac{\delta{\cal {L}}}{\delta\varphi}=\frac{\partial{\cal {L}}}{\partial\varphi}-D_{t}^{L}\left(\frac{\partial{\cal {L}}}{\partial\varphi_{t}}\right)-D_{\xi}\left(\frac{\partial{\cal {L}}}{\partial\varphi_{\xi}}\right)=0,
\end{equation}
where the operators $D_{\xi}$ and $D_{t}^{L}$ are the total derivatives
in Lagrangian coordinates.

Assume that ${\cal L}={\cal L}(\varphi,\varphi_{t},\varphi_{\xi},S_{0}(\xi))$.
Substituting this Lagrangian into the equation
\[
\frac{\delta{\cal L}}{\delta\varphi}=0,
\]
then excluding $\varphi_{t\xi}$ found from (\ref{eq:Nov25.10}),
and splitting it with respect to the derivatives $\varphi_{tt}$,
$\varphi_{\xi\xi}$ and $S_{0}^{\prime}$, one finds an overdetermined
system of partial differential equations. Solving this overdetermined
system, one derives the Lagrangian
\begin{equation}
{\cal L}=\Gamma+{\displaystyle \frac{1}{\gamma-1}\Gamma^{\gamma}S\varphi_{\xi}^{1-\gamma}.}\label{eq:Lagrangian}
\end{equation}
Thus, one obtains that equation (\ref{eq:Nov25.10}) is indeed the
Euler-Lagrange equation of the Lagrangian (\ref{eq:Lagrangian}).

\section{Group properties of equation (\ref{eq:Nov25.10})}


As mentioned in the Introduction, for purpose of using Noether's theorem
of equation (\ref{eq:Nov25.10}), one needs to study its group properties.

\subsection{Equivalence transformations}

An equivalence group allows changing arbitrary elements, yet conserving
the structure of the studied equations. An infinitesimal generator
$X$ of the equivalence group is sought in the form \cite{bk:Ovsiannikov1978}
\begin{equation}
X^{e}=\zeta^{\xi}\partial_{\xi}+\zeta^{t}\partial_{t}+\eta^{\varphi}\partial_{\varphi}+\eta^{S}\partial_{S},
\end{equation}
where the coefficients $\zeta^{\xi},\zeta^{t},\eta^{\varphi}$ and
$\eta^{s}$ are functions of all variables \cite{bk:Meleshko[2005]}:
$(\xi,t,\varphi,S)$. Here the function $S(\xi)$ is an arbitrary
element of equation (\ref{eq:Nov25.10}). Calculations give the following
basis of generators of the equivalence group
\[
\begin{array}{l}
X_{1}^{e}=\partial_{\xi},\quad X_{2}^{e}=\partial_{t},\quad X_{3}^{e}=\partial_{\varphi},\quad X_{4}^{e}=\varphi\partial_{t}+t\partial_{\varphi},\\
X_{5}^{e}=\xi\partial_{\xi}+t\partial_{t}+\varphi\partial_{\varphi},\quad X_{6}^{e}={\displaystyle \frac{\xi}{\gamma-1}\partial_{\xi}+S\partial_{S}.}
\end{array}
\]

The equivalence transformations corresponding to the generators which
can change the arbitrary element are
\[
\begin{array}{l}
X_{1}^{e}:\,\,\,\,\,\,\,\tilde{\xi}=\xi+a,\\
X_{5}^{e}:\,\,\,\,\,\,\tilde{\xi}=\xi e^{a},\tilde{t}=te^{a},\tilde{\varphi}=\varphi e^{a},\tilde{S}=S,\\
X_{6}^{e}:\,\,\,\,\,\,\tilde{\xi}=\xi e^{{\displaystyle \frac{a}{\gamma-1}}},\tilde{\epsilon}=Se^{a},
\end{array}
\]
where $a$ is the group parameter, and only changeable variables are
presented.

\subsection{Group classification of the Euler-Lagrange equation (\ref{eq:Nov25.10}) }

A symmetry $X$ of equation (\ref{eq:Nov25.10}) is sought in the
form
\begin{equation}
X=\zeta^{\xi}\partial_{\xi}+\zeta^{t}\partial_{t}+\eta^{\varphi}\partial_{\varphi},\label{eq:A1}
\end{equation}
where the unknown coefficients $\zeta^{\xi}$, $\zeta^{t}$, and $\eta^{\varphi}$
are functions of $(\xi,t,\varphi)$. Applying the prolonged generator
to equation (5), substituting the main derivative, and splitting with
respect to parametric derivatives, one derives that
\[
\zeta^{t}=k_{4}t+k_{3}\varphi+k_{2},\quad\eta^{\varphi}=k_{3}t+k_{4}\varphi+k_{1},
\]
\begin{equation}
{\displaystyle \frac{\partial\zeta^{\xi}}{\partial\xi}=-\frac{\zeta^{\xi}S_{0}^{\prime}}{S_{0}(\gamma-1)}+k_{4},}\label{eq:A6}
\end{equation}
and obtains the classifying equation
\begin{equation}
\zeta^{\xi}\left(S_{0}^{\prime\prime}S_{0}(\gamma-1)-S_{0}^{\prime}{}^{2}\gamma\right)+S_{0}^{\prime}S_{0}(\gamma-1)k_{4}=0,\label{eq:A7}
\end{equation}
where $k_{1},k_{2},k_{3}$ and $k_{4}$ are constants. A basis of
the kernel of admitted Lie algebras, which is a Lie algebra admitted
by equation (\ref{eq:Nov25.10}) for all functions $S_{0}(\xi)$,
consists of the generators
\begin{equation}
X_{1}=\partial_{\varphi},\quad X_{2}=\partial_{t},\quad X_{3}=\varphi\partial_{t}+t\partial_{\varphi}.\label{eq:A8}
\end{equation}
Further analysis of the classifying equation (\ref{eq:A7}) is similar
to the analysis represented for the non-relativistic case \cite{bk:DorodnitsynKozlovMeleshko2019}.

\subsubsection{Case $S_{0}^{\prime}=0$}

In this case $S_{0}$ is constant, and the admitted Lie algebra is
defined by the generators$X_{1}$, $X_{2}$, $X_{3}$ and
\[
X_{4}=\xi\partial_{\xi}+t\partial_{t}+\varphi\partial_{\varphi},\quad X_{5}=\partial_{\xi}.
\]

\subsubsection{Case $S_{0}^{\prime}\protect\neq0$}

From equation (\ref{eq:A7}), one obtains
\begin{equation}
k_{4}=-{\displaystyle \frac{\zeta^{\xi}}{(\gamma-1)S_{0}S_{0}^{\prime}}\left(S_{0}^{\prime\prime}S_{0}(\gamma-1)-S_{0}^{\prime}{}^{2}\gamma\right).}\label{eq:CCC}
\end{equation}
As $k_{4}$ is constant, then differentiating equation (\ref{eq:CCC})
with respect to $\xi$, one gets
\[
\zeta^{\xi}\Delta=0,
\]
where $\Delta=-S_{0}S_{0}^{\prime}S_{0}^{\prime\prime\prime}+2S_{0}S_{0}^{\prime\prime}{}^{2}-S_{0}^{\prime\prime}S_{0}^{\prime}{}^{2}$.
For existence of an extension of the kernel of admitted Lie algebras
(\ref{eq:A8}) one needs that
\begin{equation}
\Delta=0.\label{eq:Dec16.1}
\end{equation}
Noting that the latter equation can be written in the form
\[
\left(\frac{S_{0}S_{0}^{\prime\prime}}{S_{0}^{\prime}{}^{2}}\right)^{\prime}=0,
\]
one finds that
\[
S_{0}^{\prime}=C_{1}S_{0}^{C_{2}},
\]
where $C_{1}$ and $C_{2}$ are constant. Depending on $C_{2}$, the
function $S_{0}(\xi)$ can take one of two forms (up to equivalence
transformations): if $C_{2}=1$, then
\begin{equation}
S_{0}(\xi)=e^{q\xi},\label{eq:Dec17.2}
\end{equation}
otherwise
\begin{equation}
S_{0}(\xi)=\xi^{q}\label{eq:Dec17.1}
\end{equation}
Here $q\neq0$ is constant.

Substituting the function $S_{0}(\xi)$ into (\ref{eq:CCC}) and (\ref{eq:A6}),
and solving a linear ordinary differential equation for the function
$\zeta(\xi)$, one obtains the following. The extension of the kernel
of admitted Lie algebras for (\ref{eq:Dec17.2}) is given by the generator
\[
X_{4a}=\left(\gamma-1\right)\partial_{\xi}+q(t\partial_{t}+\varphi\partial_{\varphi}),
\]
and for the entropy (\ref{eq:Dec17.1}), it is
\[
X_{4b}=(\gamma-1)\xi\partial_{\xi}+(\gamma+q-1)(t\partial_{t}+\varphi\partial_{\varphi}).
\]

\textbf{Remark}. In Eulerian coordinates, the solutions corresponding
to relation (\ref{eq:Dec17.1}) are described by the differential
constraint
\begin{equation}
q\rho SS_{xx}+(1-q)\rho S_{x}^{2}-q\rho_{x}SS_{x}=0.\label{eq:dif_con1}
\end{equation}
The assumption for the entropy (\ref{eq:Dec17.2}) is given in Eulerian
coordinates by the differential constraint
\begin{equation}
S_{x}=\rho qS.\label{eq:dif_con2}
\end{equation}
One can check that the overdetermined systems of equations consisting
of the gas dynamics equations and these constraints are involutive.

\section{Conservation laws}

This section deals with conservation laws obtained by using Noether's
theorem. Noether's condition 
has the following form
\begin{equation}
X\mathcal{L}+\mathcal{L}(D_{t}^{L}\zeta^{t}+D_{\xi}\zeta^{\xi})=D_{t}^{L}B_{1}+D_{\xi}B_{2}.\label{eq:BBB}
\end{equation}
for some functions $B_{i}(t,\xi,\varphi)$, $i=1,2$. Here we assume
that the operator $X$ is prolonged to the derivatives by means of
the standard prolongation formulas~\cite{bk:Ovsiannikov1978,bk:Olver[1986]}.
The symmetry $X$ is called variational if $B_{1}=B_{2}\equiv0$;
for divergence symmetries there are nontrivial $B_{1}$ and $B_{2}$.

The densities $(T^{t},T^{\xi})$ of the conservation laws are given
by the formulae
\begin{equation}
\begin{array}{c}
T^{t}=\zeta^{t}{\cal L}+(\eta^{\varphi}-\zeta^{t}\varphi_{t}-\zeta^{\xi}\varphi_{\xi})\frac{\partial{\cal L}}{\partial\varphi_{t}}-B_{1},\\[1.5ex]
T^{\xi}=\zeta^{\xi}{\cal L}+(\eta^{\varphi}-\zeta^{t}\varphi_{t}-\zeta^{\xi}\varphi_{\xi})\frac{\partial{\cal L}}{\partial\varphi_{\xi}}-B_{2}.
\end{array}\label{conservation_laws_densities}
\end{equation}

\subsection{Conservation laws corresponding to the kernel}

All symmetries corresponding to the kernel of admitted Lie algebras
are variational. Using Noether's theorem, the conservation laws are
\[
T_{1}^{\xi}=\Gamma^{\gamma}S_{0}\varphi_{\xi}^{-\gamma},\,\,\,T_{1}^{t}=\varphi_{t}\Gamma^{-1}G;
\]
\[
T_{2}^{\xi}=S_{0}\Gamma^{\gamma}\varphi_{t}\varphi_{\xi}^{-\gamma},
\]
\[
T_{2}^{t}=\Gamma^{-1}G-S_{0}\Gamma^{\gamma}\varphi_{\xi}^{1-\gamma};
\]
\[
T_{3}^{\xi}=S_{0}\Gamma^{\gamma}(\varphi\varphi_{t}-t)\varphi_{\xi}^{-\gamma},
\]
\[
T_{3}^{t}=\varphi\Gamma^{-1}(1-S_{0}\varphi_{\xi}^{1-\gamma}\Gamma^{\gamma-2})-t\varphi_{t}G.
\]
where $G=\left(1+{\displaystyle \frac{\gamma S_{0}\varphi_{\xi}^{1-\gamma}\Gamma^{\gamma-1}}{\gamma-1}}\right)$.

In Eulerian coordinates, these conservation laws become
\[
T_{1}^{x}=\Gamma^{\gamma}Sm^{\gamma},\,\,\,{}^{e}T_{1}^{t}=v\Gamma^{-1}G^{E};
\]
\[
T_{2}^{x}=S\Gamma^{\gamma}vm^{\gamma},
\]
\[
^{e}T_{2}^{t}=\Gamma^{-1}G^{E}-S\Gamma^{\gamma}m^{\gamma-1};
\]
\[
T_{3}^{x}=S\Gamma^{\gamma}(xv-t)m^{\gamma},
\]
\[
^{e}T_{3}^{t}=x\Gamma^{-1}(1-Sm^{\gamma-1}\Gamma^{\gamma-2})-tvG^{E}.
\]
where $G^{E}=\left(1+{\displaystyle \frac{\gamma Sm^{\gamma-1}\Gamma^{\gamma-1}}{\gamma-1}}\right)$.

The conservation laws of momentum and energy correspond to $(^{e}T_{1}^{t},T_{1}^{x})$
and $(^{e}T_{2}^{t},T_{2}^{x})$. The conservation law with $(^{e}T_{3}^{t},T_{3}^{x})$
is similar to the conservation law of center of mass in non-relativistic
gas dynamics.

\subsection{Conservation laws for constant entropy}

The generator $X_{4}$ does not satisfies condition (\ref{eq:BBB}).
The generator $X_{5}=\partial_{\xi}$ provides the conservation law
with
\[
\begin{array}{l}
T_{5}^{\xi}=\Gamma G{\displaystyle ,\ T_{5}^{t}={\displaystyle \varphi_{t}\varphi_{\xi}\Gamma^{-1}G.}}\end{array}
\]
The densities of this conservation law in Eulerian coordinates are
\[
T_{5}^{x}={\displaystyle \Gamma G^{E},}\,\,\,{}^{e}T_{5}^{t}=v\Gamma^{-1}m^{-1}G^{E}.
\]

\subsection{Conservation laws for a particular entropy}

The generator $X_{4a}$ does not satisfies condition (\ref{eq:BBB}).

Substituting the generator $X_{4b}$ into equation (\ref{eq:BBB}),
one obtains
\[
k_{4}(2(\gamma-1)+q)=0.
\]
Hence, an additional conservation law exists if $q=2(1-\gamma)$.
This conservation law has the densities
\[
T_{4}^{\xi}=\xi\Gamma G+S_{0}\Gamma^{\gamma}\varphi_{\xi}^{-\gamma}(\varphi-t\varphi_{t}),
\]
\[
T_{4}^{t}=t(S_{0}\Gamma^{\gamma}\varphi_{\xi}^{1-\gamma}-\Gamma^{-1}G)+(\varphi+\xi\varphi_{\xi})\varphi_{t}\Gamma^{-1}G.
\]
In Eulerian coordinates, this conservation law becomes
\[
T_{4}^{x}=S^{1/q}\Gamma G^{E}+S\Gamma^{\gamma}m^{\gamma}(x-tv),
\]
\[
^{e}T_{4}^{t}=t(S\Gamma^{\gamma}m^{\gamma-1}-\Gamma^{-1}G^{E})+(x+S^{1/q}m^{-1})v\Gamma^{-1}G^{E}.
\]

\section{Conclusions}

The present paper is devoted to the analysis of the one-dimensional
relativistic gas dynamics equations. The studied equations are considered
in Lagrangian description. This description allows us to find a Lagrangian
such that the relativistic gas dynamics equations can be rewritten
in a variational form. One of the advantages of the variational form
is that it makes an application of Noether's theorem for deriving
conservation laws possible. The present paper is focused on application
of Noether's theorem for deriving conservation laws in Lagrangian
description. Application of Noether's theorem requires the group analysis
of the Euler-Lagrange equation. In the present paper we performed
the complete group classification of the Euler-Lagrange equation with
respect to the entropy. This classification of symmetries of the Euler-Lagrange
equation allowed us to derive additional conservation laws beyond
the classical ones. The analogs of the found conservation laws in
Eulerian coordinates are also presented.

\section*{Acknowledgements}
The research  was supported by Russian Science Foundation Grant No 18-11-00238
`Hy\-dro\-dynamics-type equations: symmetries, conservation laws, invariant difference schemes'.
The authors thank E.Shulz for the assistance.


\end{document}